\documentclass[aps,prl,superscriptaddress,twocolumn]{revtex4}
\usepackage{amssymb}
\usepackage{graphicx}
\usepackage{amsmath}
\usepackage{color}
\usepackage{bm}
\usepackage{ulem}

\begin{document}

\title{Machine Learning Identification of Impurities in the STM Images}

\author{Ce Wang}
\affiliation{Institute for Advanced Study, Tsinghua University, Beijing, 100084, China}

\author{Haiwei Li}
\affiliation{State Key Laboratory of Low Dimensional Quantum Physics, Department of Physics, Tsinghua University, Beijing 100084, China}

\author{Zhenqi Hao}
\affiliation{State Key Laboratory of Low Dimensional Quantum Physics, Department of Physics, Tsinghua University, Beijing 100084, China}

\author{Xintong Li}
\affiliation{State Key Laboratory of Low Dimensional Quantum Physics, Department of Physics, Tsinghua University, Beijing 100084, China}

\author{Cangwei Zou}
\affiliation{State Key Laboratory of Low Dimensional Quantum Physics, Department of Physics, Tsinghua University, Beijing 100084, China}

\author{Peng Cai}
\email{pcai@ruc.edu.cn}
\affiliation{Department of Physics and Beijing Key Laboratory of Opto-electronic Functional Materials and Micro-nano Devices, Renmin University of China, Beijing, 100872, China}

\author{Yayu Wang}
\affiliation{State Key Laboratory of Low Dimensional Quantum Physics, Department of Physics, Tsinghua University, Beijing 100084, China}
\affiliation{Frontier Science Center for Quantum Information, Beijing 100084, China}

\author{Yi-Zhuang You}
\email{yzyou@ucsd.edu}
\affiliation{Department of Physics, University of California, San Diego, California 92093, USA}

\author{Hui Zhai}
\email{hzhai@tsinghua.edu.cn}
\affiliation{Institute for Advanced Study, Tsinghua University, Beijing, 100084, China}
\date{\today}

\begin{abstract}
In this work we train a neural network to identify impurities in the experimental images obtained by the scanning tunneling microscope measurements. The neural network is first trained with large number of simulated data and then the trained neural network is applied to identify a set of experimental images taken at different voltages. We use the convolutional neural network to extract features from the  images and also implement the attention mechanism to capture the correlations between images taken at different voltages. We note that the simulated data can capture the universal Friedel oscillation but cannot properly describe the non-universal physics short-range physics nearby an impurity, as well as noises in the experimental data. And we emphasize that the key of this approach is to properly deal these differences between simulated data and experimental data. Here we show that even by including uncorrelated white noises in the simulated data, the performance of neural network on experimental data can be significantly improved. To prevent the neural network from learning unphysical short-range physics, we also develop another method to evaluate the confidence of the neural network prediction on experimental data and to add this confidence measure into the loss function. We show that adding such an extra loss function can also improve the performance on experimental data. Our research can inspire future similar applications of machine learning on experimental data analysis.     
\end{abstract}

\maketitle

\textit{Introduction.} Machine learning algorithms are good at extracting features from patterns, which have found broad applications in industry such as face recognition and imaging processing. In most of physics experiments, data analysis is essentially a task of pattern recognization, and one needs to extract useful information from images taken from experiments. Therefore, recently there is a general trend of applying machine learning algorithms to data analysis in physics experiments \cite{ziatdinov2016,bohrdt2019,rem2019,zhang2019,samarakoon2019,Torlai2019,khatami2020}. The success of this approach not only can make the data analysis processes more efficiently, but also can possibly reveal hidden features and new physics behind the experimental images that are hard to be identified by human eyes.

There are different kinds applications of pattern recognization algorithms on experimental data.  One typical application is clustering data with common features by unsupervised learning methods and extracting the essential common features by imposing an information bottleneck \cite{Bishop}. This usually works for situations where images are abundant. In this work we focus on another type of applications. In these cases, the experimental data are not abundant, but one has idea from theory about what kinds of features that we are looking for. In such cases, one can first train the neural network (NN) with big data generated by theoretical simulation of the features, and then apply the trained NN to make predictions on the experimental data. For instances, one can train a NN to learn simulated STM images with charge-density-wave of different wave vectors, and then the trained NN is applied to real STM data to recognize whether these data contain feature of any of these kinds of charge-density-wave \cite{zhang2019}. One can also train a NN to learn the real-space single-site density distribution of the Fermi Hubbard model generated by different approximated theories, and then the trained NN is applied to quantum gas microscope data to tell which theory fits the data better \cite{bohrdt2019}.  

However, while we need the simulation data to train the NN, we also need to note the differences between the simulation and the reality. The differences are mainly two folds. First, in real systems there inevitably exist noises. In many cases, it is hard to quantify the degree of noises in reality. Secondly, most theoretical simulations can only capture the universal part of a feature and cannot capture the non-universal part. Usually, the long-range and low-energy part is universal, but the short-range and high-energy part depends on various details, which varies between different samples. We will elaborate these differences in detail in the example that we will discuss in this work. Therefore, we should prevent the NN from overfitting the simulation data, otherwise the trained model will not generalize well to realistic cases. In this work, we propose two approaches to avoid overfitting these unrealistic features in the simulation data, that are the \textit{data augmentation} and the \textit{confidence regularization}, respectively. One of the key point of this work is to show that, when a NN is trained with the simulated data, whether and how one mitigates the overfitting of unrealistic features can significantly change the outcome when the trained NN is applied to experimental data. Therefore, one should be very careful when using this stratagem to judge a theory \cite{bohrdt2019}. We will demonstrate this point by a concrete task, that is to identify the locations of impurities from images obtained by the scanning tunneling microscope (STM). 

\begin{figure}[t]
\begin{center}
\includegraphics[width=0.48\textwidth]{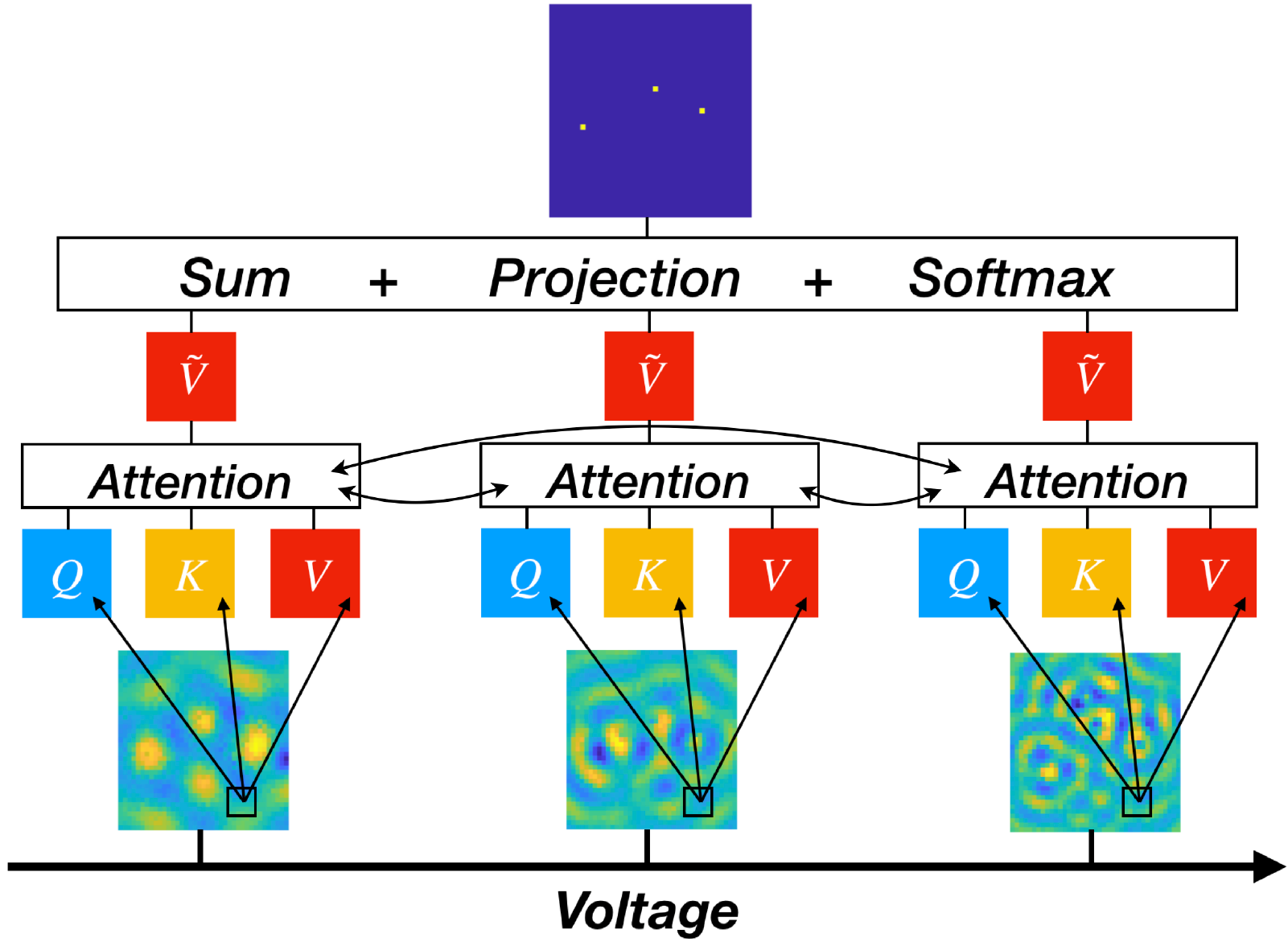}
\end{center}
\caption{The structure of the NN. The voltage arrow indicates a sequence of images taken at different voltages. Each image $X^l_i$ is mapped to tensors of the query $Q^l_i$, the key $K^l_i$ and the value $V^l_i$. By the attention mechanism, it leads to normalized value $\tilde{V}^l_i$ computed by Eq. \ref{attention_eq}. Then, we sum over all $\tilde{V}^l_i$ and reach the output following a linear projection and a Softmax.   }
\label{network}
\end{figure}

\textit{Identification of Impurities in the STM Imaging.} The STM measurements result in two-dimensional maps of the local electron density-of-state (DoS) at the surface of a material. For a given surface of a sample, by varying the voltage bias, one can obtain a sequences of maps, which correspond the local DoS at different Fermi energies. Usually there are always impurities in the materials, and these impurities can manifest themselves in the local electron DoS. Our goal is to identify the locations of impurities from these images. Therefore, we would like to build a NN to learn the mapping from the STM images ${\bf X}^l$ to the distribution of impurities $Y^l$, where $l$ labels samples in the data set. More concretely, each input of the NN ${\bm X}^{l}$ contains a sequence of two-dimensional images as $\{X^l_i,\{i=1,\dots,N_v\}\}$, where $i$ labels images on the same surface but with different Fermi energies. $N_v$ is the total number images measured at a given surface. $Y^l$ is a two-dimensional probability distribution function, which we normalized to unity.

Here we should emphasize that different $X_i^l$ with the same $l$ are actually correlated, because if the presence of one impurity manifests itself in the locally DoS with one voltage, the similar features must consistently appear in images with other voltages. On the other hand, if similar feature occurring in one image comes from noises, it will not show up in images with other voltages. That is to say, if we properly capture the correlations between images with different voltages, it can help us detect impurities more accurately.  

\begin{figure}[t]
\begin{center}
\includegraphics[width=0.85\columnwidth]{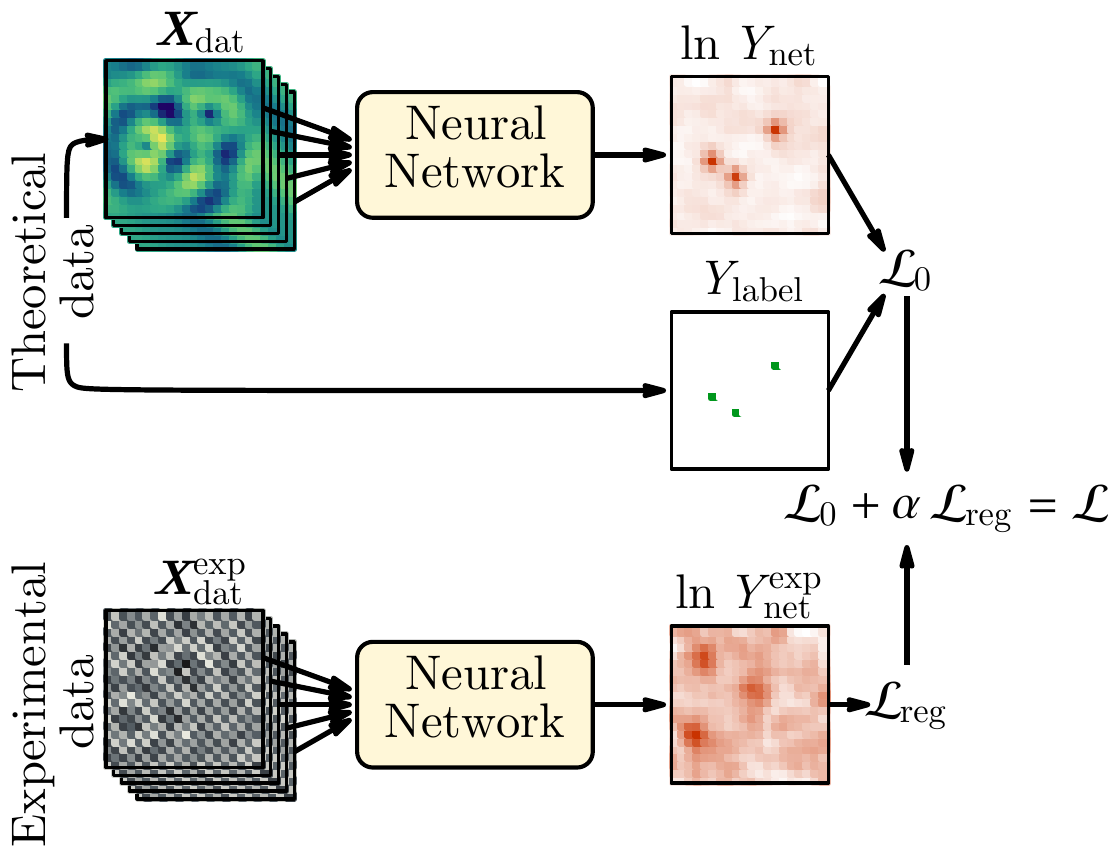}
\caption{Computation graph for the loss function. The NN is trained on the theoretical data via the regression loss $\mathcal{L}_0$, and regularized by the experimental data via the prediction confidence regularization $\mathcal{L}_\text{reg}$.}
\label{fig: loss}
\end{center}
\end{figure}

\textit{Design of the NN.}  To capture the correlation between images taken at different voltages, we implement \textit{attention mechanism} in our NN. Attention mechanism has been widely used in nature language processing \cite{attention,bert}, which can capture the long distance correlation between different sentences. Here we treat the STM images from different voltages as different ``sentences'', and apply the self-attention approach to aggregate the consistent features among these images. More concretely, our NN structure is shown in Fig. \ref{network}. As mentioned above, each input data ${\bm X}^l$ is made of a sequence of images $X^l_i$. Here we consider each $X^l_i$ as a $L\times L$ pattern. Here we first apply the convolutional NN to general three tensors $Q^l_i$, $K^l_i$ and $V^l_i$. The convolutional NN is commonly used for extracting features from images. In term of the terminology of attention, they are called the query, the key and the value, respectively. Their dimensions are $L\times L\times d_q$, $L\times L\times d_k$ and $L\times L\times d_v$, respectively. Following the scheme of attention, we introduce the normalized value $\tilde{V}^l_i$ as
\begin{equation}
\tilde{V}^l_i=\frac{1}{Z^l_i}\sum_jV^l_j\exp(\langle Q^{l}_i K^l_j\rangle). \label{attention_eq}
\end{equation}   
where $\langle Q^{l}_i K^l_j\rangle$ denotes the inner-product between the query and the key. For taking the inner-product, we need $d_q=d_k$, and $Z^l_i=\sum_j\exp(\langle Q^l_i K^l_j\rangle)$. After the attention, we sum over all $\tilde{V}^l_i$ and obtain $V^l_\text{T}=\sum_{i}\tilde{V}^l_i$. We apply a linear projection to project $V^l_\text{T}$ from $L\times L\times d_{v}$ onto $L\times L$ dimensional $\mathcal{S}^l$. Finally, by applying a Softmax to $\mathcal{S}$, we obtain a $L\times L$ dimensional image $Y^l_\text{net}$ as the output of the NN. The Softmax function automatically ensures that $Y^l_\text{net}$ is normalized to unity. This output should be compared with the labelled impurity distribution $Y^{l}_\text{label}$. The loss function can be defined as the cross entropy between $Y^l_\text{net}$ and $Y^l_\text{label}$
\begin{equation}
\mathcal{L}_{0} = -\sum\limits_{l}Y^{l}_{\text{label}}\odot\ln Y^{l}_{\text{net}}, 
\end{equation} 
where the operator $\odot$ means an element-wise multiply followed by a summation over all matrix elements. Minimizing the cross entropy $\mathcal{L}_0$ aims to bring the predicted impurity distribution $Y^l_\text{net}$ close to the ground truth $Y^l_\text{label}$ as much as possible. The NN can be trained by the standard gradient decent method. 

\textit{Training Date Generation.} Here we describe how we prepare the training data according to the theoretical knowledge about impurities. Suppose we have $N_i$ impurities in an area discretized into $L\times L$ pixels, and their locations are ${\bm r}_w$ ($w=1,\dots,N_i$). The presence of an impurity induces the Friedel oscillation in a metal, and all these impurities lead to a density modulation as
\begin{equation}
\delta n({\bf r})=\sum\limits_{w=1}^{N_i}A_w\frac{\cos(2k_\text{f} |{\bf r}-{\bf r}_w|+\alpha_w)}{(|{\bf r}-{\bf r}_w|+\epsilon)^2}.  \label{Frid}
\end{equation} 
We have taken the spacing between pixels as unity, such that all the other quantities are also made dimensionless accordingly. Each set of impurity position $\{ {\bf r}_w, w=1,\dots, N_i\}$ gives rise to a fixed impurity configuration, which is labelled by $l$. By varying the impurity position $\{{\bf r}_w\}$, we can easily generate a large number of data. We estimate the charge carrier density of the material, from which we can determine a typical value $k^0_\text{f}$. For a given impurity configuration, we sample a number of different $k_\text{F}$ around $k^0_\text{F}$, which correspond to measurements at different voltages labelled by $i$. For a given impurity configuration and a given $k_\text{F}$, we randomly choose parameters $\{A_w\}$ and $\{\alpha_w\}$. $A_w$ is governed by the strength of the impurity, and $\alpha_w$ depends on the short-range condition of the impurity.  For each $\delta n^l_i({\bf r})$, we calculate $\bar{n}^l_i=\int \delta n^l_i({\bf r})$, and by subtracting $\bar{n}^l_i$, we can make $\delta n^l_i({\bf r})$ averaged to zero. Moreover, we normalize the amplitude of $\delta n^l_i$ such that $\delta n^l_i$ always lies between $[-1,1]$. This $\delta n^l_x$ is take as input $X^l_i$. In this way, we generate the input ${\bm X}^l$. For the output, if there is an impurity located at one pixel, we assign this pixel a probability $1/N_i$. Otherwise, we assign it zero probability. In this way, we generate $Y^l$ as output. 

Here we would like to highlight the difference between these generated data and the reality. First, there is no noise in these data. We propose to add a white noise in each pixel which obeys a Gaussian distribution with width $\eta$. Although in reality the noises are not  uncorrelated white noises and it is hard to know the actual noise distribution in reality, adding uncorrelated white noise is the most natural choices. Below we will compare the prediction power of the NN trained by data with or without noises. Secondly, we should point out that, strictly speaking, the universal form of the Friedel oscillation Eq. \ref{Frid} is only valid at a distance $\gtrsim 1/k_\text{F}$ away from an impurity. Close to an impurity, the density distribution strongly depends on the detailed potential form of the impurity. Since we do not have prior knowledge of the impurity potential, we cannot simulate accurately the short-range part of the impurity potential. In Eq. \ref{Frid} we simply add a cutoff $\epsilon$ to avoid the divergence when ${\bf r}\rightarrow {\bf r}_w$. Therefore, the simulated data cannot describe the short-range physics nearby an impurity properly. Hence, if the NN overfits the short-range features of the training data, it can actually hinder the preference of the NN on actual experimental data. We will also discuss how to deal with this issue below.
   
\begin{figure}[t]
\begin{center}
\includegraphics[width=0.45\textwidth]{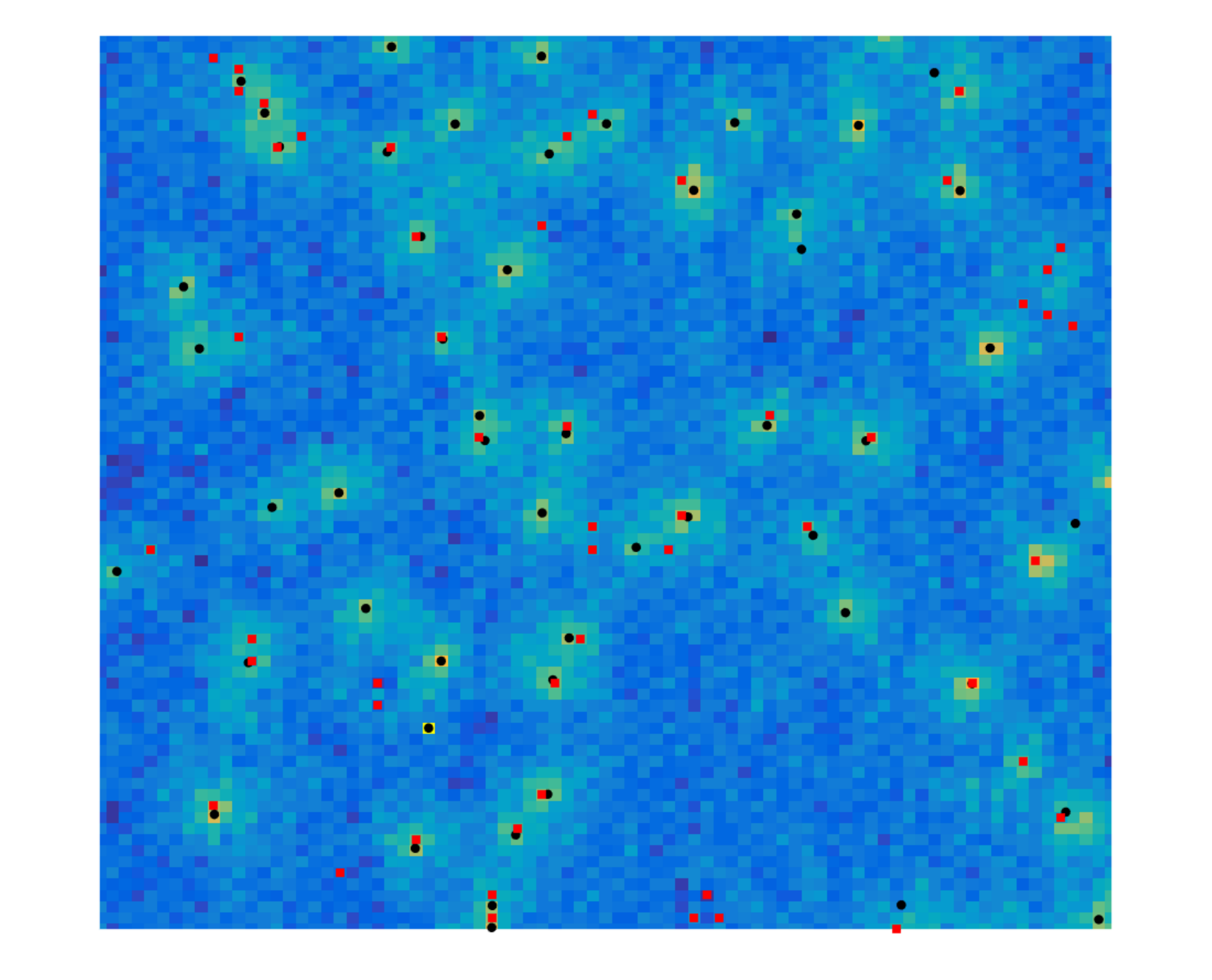}
\end{center}
\caption{Comparison between the prediction of the NN and a reference answer from another experiment. Red squares are the prediction of NN, determined by local maxima of the $\mathcal{S}$ layer generated by the NN. Black dots are the reference answer from another experiment, determined by the local maxima of the DoS of bound states shown as the color plot. Here, for training the NN, we use loss function $\mathcal{L}_0+\alpha \mathcal{L}_\text{reg}$ with $\alpha=0.03$ and training data with $\eta=0.8$ Gaussian noise.  }
\label{result}
\end{figure}

\textit{Results.} Here we train the NN by a large number of simulated data with $L=20$. The experimental data is taken at one sample of the Gold surface, which contains a sequence of images taken at difference bias voltages. The experimental details can be found in the appendix. We work on an area of $80\times 80$ large, which are further divided on sixteen $20\times 20$ images. We apply the NN to examine each of the $20\times 20$ images individually. Then, we pack the predictions together and compare with a reference measurement. To properly compare different images, instead of plotting $Y$, we consider the layer $\mathcal{S}$ before the Softmax function. We take the local maxima of $\mathcal{S}$ as the locations of impurities. The result is shown in Fig. \ref{result}.

 \begin{figure}[t]
\begin{center}
\includegraphics[width=0.48\textwidth]{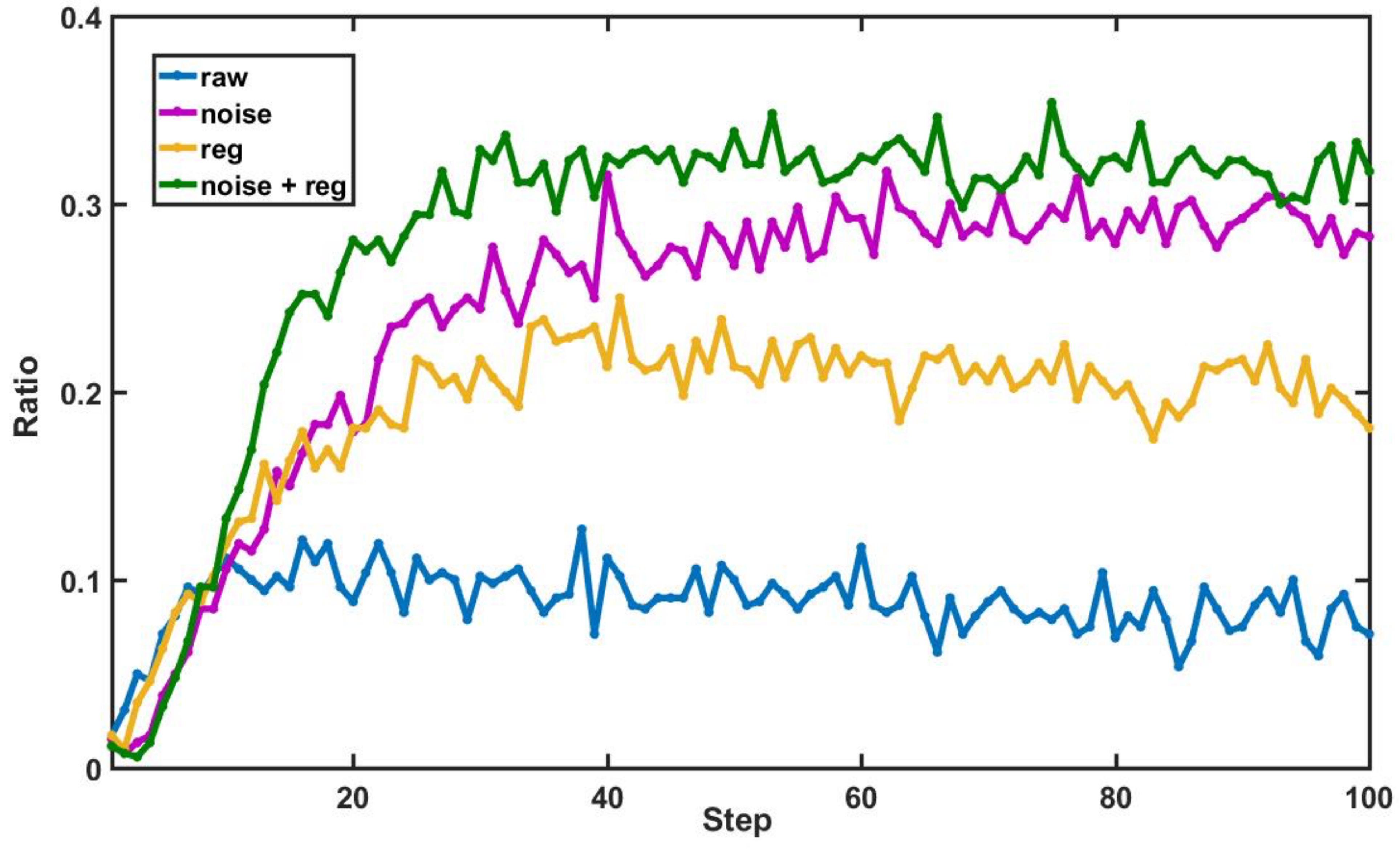}
\end{center}
\caption{Compare the performance of different approaches on the experimental data. The curves are the ratio of the reference points perfectly agreed with prediction to the total reference points.  The blue line and the purple line are given by NN trained by loss function $\mathcal{L}_0$, and the training data have no noise for the blue line and include noise for the purple line. The yellow line is given by NN trained by loss function $\mathcal{L}_0+\alpha \mathcal{L}_\text{reg}$ including confidence regularization and data without noise. Finally, the green line is given by NN trained by loss function $\mathcal{L}_0+\alpha \mathcal{L}_\text{reg}$ and data with noise.    Here we take $\alpha=0.03$. The ratios are plotted as a function of the training epoch, and the results are averaged over ten independent training processes.     }
\label{compare}
\end{figure}

Here let us briefly discuss the reference measurement, and we also refer reader to appendix for more details. As we have discussed above, the experimental images to be recognized are taken for the local DoS around the Fermi energy. On the other hand, we can also take the experimental data at another voltage far away from the Fermi energy, which actually measures the bound state due to the impurity potential. Therefore, this measurement essentially reveals the amplitude of the bound state wave function. We use expectation-maximization algorithm \cite{Bishop} to decide the positions of impurities for the bound state measurement and the results are taken as reference answer. As shown in Fig. \ref{result}, red squares are impurity locations predicted by the trained NN and black points are impurity locations given by the reference measurement. Though our NN has absolutely no access to the bound state data, we can see that some predicted peaks perfectly coincide with the reference answer. We count the reference points whose distance from the nearest predicted peak is less than one pixel, which means nearly perfect agreements, and we determine the ratio of such points to the total reference points. We use this ratio to measure the preference of the NN.

\underline{Effect of Data Augmentation:} In Fig. \ref{compare} we plot the ratios as a function of the training epoch of the NN. We can see that the ratios increase as the training epochs increase. The blue line is given by NN trained by data without noises and the purple line is given by NN trained by data with noises.  It is very clear that the performance of the NN can be significantly improved by including noises in the training data. Here we have taken $\eta=0.8$ for noise distribution. 

\underline{Effect of Confidence Regularization:} As we have mentioned above, another difference between the simulated data and the reality is the short-range physics. When the NN begins to learn the short-range details, it experiences frustration when applied to the real data, and consequently, the predictions on the real data become less confident. Here we invent a new scheme to deal with this problem. The essential idea is to force the NN to make confident predications on the real data.  At each training step, we can apply the NN to the real experimental data and the NN returns an output, denoted by $Y_\text{net}^{\text{exp}(m)}$, where $m$ labels different experiment regions. There could be different ways to define and evaluate the confidence level of the prediction. A natural choice is to look at the entropy of the predicted impurity distribution 
\begin{equation}
\mathcal{L}_\text{reg} = - \sum \limits_{m} Y_{\text{net}}^{\text{exp}(m)} \odot \ln Y_{\text{net}}^{\text{exp}(m)}.
\end{equation}
If the NN can make confident prediction on the real data, the predicted distribution will be sharply concentrated at a few positions. A sharper distribution will be more informative and hence has less entropy. We introduce $\mathcal{L}_\text{reg}$ as an confidence regularization to our loss function, and write the full loss function $\mathcal{L}$ as $\mathcal{L}_0 + \alpha \mathcal{L}_\text{reg}$, as illustrated in Fig.~\ref{fig: loss}. When we minimize this loss function, it minimizes $\mathcal{L}_0$, which drives the NN to learn from the simulated training data, and simultaneously,  it also minimizes $\mathcal{L}_\text{reg}$, which encourages the NN to make sharp predictions on the real data. In this way, the real data is also included to regularize the training process.  Here $\alpha$ controls the strength of the regularization. In Fig. \ref{compare}, $\alpha=0.03$ is taken to achieve best performance. It is also clear that the performance improves significantly by including the confidence regularization. The best performance is achieved with both proper confidence regularization and adding noises in simulated data.                 
 
\textit{Concluding Remarks.} In physics research, it quite often that we need to verify certain theory with experimental data. With the help of the machine learning algorithm, what we can do is to train the NN to learn data generated by theory, and then applied the trained NN to detect whether certain features exist in the experimental data. The success or failure in detecting this feature can be viewed as a support or a falsification of the corresponding theory. This can become a promising canonical method in future physics research. Here we point out that how to properly treat the non-universal part of the feature and the presence of noise in reality can be crucial in this approach. The attempts of this work can be quite useful in future developments of this approach.   
 
\textit{Acknowledgment.} HZ is supported by Beijing Outstanding Scholar Program, MOST under Grant No. 2016YFA0301600 and NSFC Grant No. 11734010. YZY is supported by a startup fund from UCSD. PC is supported by the Fundamental Research Funds for the Central Universities, and the Research Funds of Renmin University of China.

 \begin{figure}[t]
\begin{center}
\includegraphics[width=0.48\textwidth]{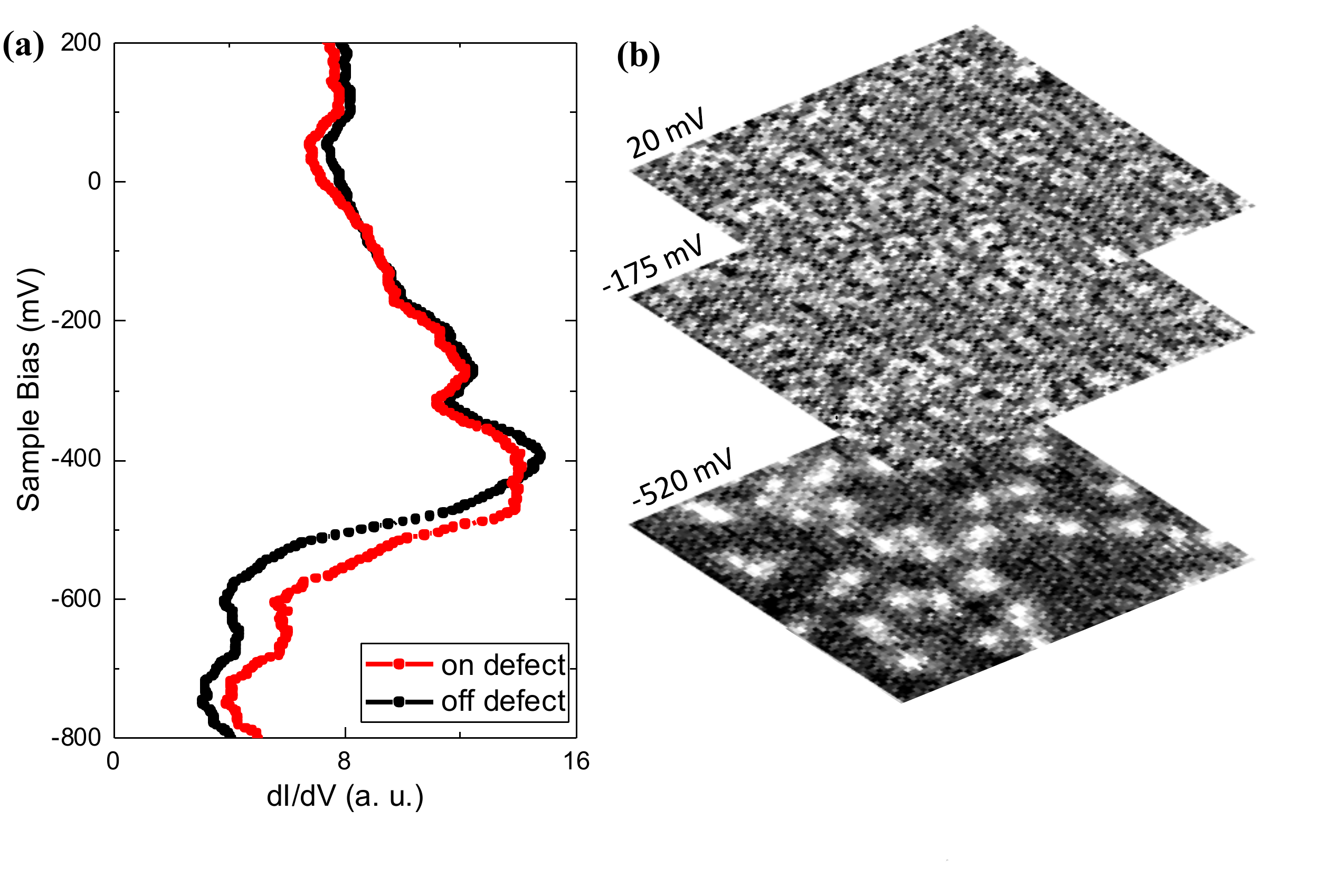}
\end{center}
\caption{ Experimental data on Au(111) surface. (a) Typical $dI/dV$ spectra measured on defect (red curve) and on defect-free area (black curve). The abrupt increase of spectra above $-520$mV shows the itinerant surface states on Au(111). Additional spectra on black curve indicates the bound state of defect below the surface band. (b)  Differential conductance maps. Top panel, two representative $dI/dV$ maps (at $-20$mV and $-175$mV) taken at $5$K with tunneling junction $R=5$Gohm, showing spatial oscillation from the quantum interference of surface states. Bottom, the conductance map at $-520$mV, revealing the spatial distribution of defect states. 
 }
\label{exp}
\end{figure}

\begin{appendix}

\textit{Appendix: Experimental Details.} Here our experimental data are based on the quantum interference of Au(111) surface states. Fig. \ref{exp}(a) shows typical spectra measured on the surface. A sharp increase of $dI/dV$ above $V=-520$mV corresponds to the appearance of two-dimensional surface state. The dispersion of the surface band is parabolic with an effective mass ration of $0.26$ \cite{Exp}. The data feed to NN are a series of differential conductance maps $dI/dV(r, eV)$ taken over $375\r{A}\times375 \r{A}$ area, where $r$ is position and $V$ is sample bias voltage. The coherent quantum interference is dominant near Fermi level, which can be evidenced by strong spatial oscillation in $dI/dV$ maps. To minimize the error to the output error from the data noise, our experimental dataset in this work focuses on the energies near Fermi level, ranging from $-230$ meV to $+20$meV, with equal spacing of $2$meV, as shown in Fig. \ref{exp}(b). These data are used for recognition by NN. Alternatively, we turn to extract the scattering centers by imaging the defect state distribution in the $dI/dV$ map at $V=-520$meV, since there are localized bound states on defects below the itinerant surface band, as shown in Fig.\ref{exp}(a). 

\end{appendix}

\end{document}